\documentclass[useAMS,usenatbib]{mn2e}

\usepackage{graphicx}
\usepackage{bm}
\usepackage{psfrag}
\usepackage{mathrsfs}
\usepackage{amssymb,amsmath}
\newcommand{\be}{\begin{eqnarray} }
\newcommand{\ee}{ \end{eqnarray} }

\newcommand{\nodata}{...}

\renewcommand{\footnotesize}{\normalsize}

\title[Long-term timing of PSR~B1259$-$63]{The kinematics and orbital dynamics of the PSR~B1259$-$63/LS~2883 system from 23 years of pulsar timing}

\author[R.~M.~Shannon, S.~Johnston and R.~N. Manchester]{ R.~M.~Shannon\thanks{E-mail: ryan.shannon@csiro.au}, S.~Johnston and R.~N.~Manchester \\
CSIRO Astronomy and Space Science, Australia Telescope National Facility, Box 76 Epping, NSW, 1710, Australia }

\begin{document}

\date{Accepted 2013 October 31;  Received 2013 October 31;  In original form 2013 August 19}

\pagerange{1--10} \pubyear{2013}

\maketitle

\begin{abstract}
We present an analysis of $23$~yr of  pulse arrival times for PSR~B1259$-$63.  The pulsar is in a binary orbit about its approximately $20 M_{\sun}$ companion LS~2883.  Our best-fitting timing solution has none of the pulse-number ambiguities that have plagued previous attempts to model the binary orbit.   We measure significant first and second time derivatives of the projected semimajor axis  and longitude of the periastron of the orbit.  These variations are found to be consistent with the precession of the orbital plane due to classical spin-orbit coupling.  The derived moment of inertia and spin of the companion are consistent with the companion rotating at near-breakup velocity.
  The system configuration is also consistent with the geometry derived from both the polarisation of the radio emission and the eclipse of the pulsar by the equatorial disc of the companion.  We find strong evidence for orbital period decay that can be attributed to mass loss from the companion star.   We also measure a significant proper motion that locates the birth of the system in the Centaurus OB1 association.   
  By combining proper motion of the pulsar with radial velocity measurements of the companion, we measure the three-dimensional velocity of the system.  This velocity is used to constrain the masses of the stars prior to the supernova explosion and the kick the pulsar received at or immediately after the explosion.

\end{abstract}

\begin{keywords}
binaries: general -- stars: kinematics and dynamics  -- pulsars: general -- pulsars:~specific (PSR~B1259$-$63) 
\end{keywords}

\section{Introduction}

The PSR~B1259$-$63/LS~2883 binary system has beguiled astronomers since its discovery in 1990 \cite[][]{1992ApJ...387L..37J}.  
The pulsar has a spin period of $\sim48$~ms and an age of $0.3$~Gyr inferred from measurements of the pulsar's spin period and period derivative.    
   It  is in a highly eccentric ($e\sim0.87$) orbit with period $P_b \sim 1237$~d about a high-mass $\gtrsim 10 M_{\sun}$ main sequence stellar companion, LS~2883.   The companion is a late Oe-type or early Be-type star that possesses a significant stellar wind and an equatorial excretion disc \cite[][]{1994MNRAS.268..430J}.    Close to periastron, the pulsar is eclipsed by the disc for $\approx100$~d at which point pulsed emission becomes undetectable.  However, the system emits continuum radio and high-energy emission, caused by interactions between the pulsar and stellar winds \cite[][]{1996MNRAS.279.1026J,2011ApJ...736L..11A}.

If the companion is aspherical and the orbital axis is misaligned from the spin axis of the companion,
there is precession of the orbital plane of the pulsar, due to coupling between the quadrupolar gravitational field of the companion and the orbit of the system.  
The precession  manifests itself as linear-in-time variations in the projected size of the orbit and position of periastron passage. The largest effects are the  change of the projected size of the orbit $\dot{x}$ and longitude of periastron $\dot{\omega}$, but  detection of higher-order derivatives of both $x$ and $\omega$  is possible.  

 This coupling is expected for this system for two reasons.  
 Firstly, the companion is expected to be oblate since, as a result of mass transfer, it is rotating at near breakup velocity \cite[][]{1984JApA....5..209V}. 
Secondly,  misalignment of the spin and orbital axis is expected if the neutron star receives a kick  at the time of the supernova explosion. 
This spin-orbit coupling has been observed in another neutron star system \cite[PSR~J0045$-$7319,][]{1996Natur.381..584K}.

If the coupling is measured, it is possible to place significant constraints on the geometry of the system \cite[][]{1996Natur.381..584K,1998MNRAS.298...67W}.     The strength of the coupling can be used infer the internal structure of the companion star through measurement of its apsidal-motion constant.  The misalignment of the spin and orbital axes can be used to determine the kick imparted on the pulsar during (or immediately after) the supernova explosion.

There have been a number of attempts to fully characterise the orbit of the PSR~B1256$-$63/LS~2883 system.   Two tasks have made measurement of spin-orbit coupling difficult.  Firstly, because of the highly eccentric orbit, the eclipse covers the large fraction of orbital phase during which the effects of coupling on the measured arrival times are the largest.   
Even when the pulsar is detectable close to periastron passage, the observed radiation is affected by propagation through the disc and stellar wind, causing both dispersive \cite[][]{1995MNRAS.275..381M}  and multi-path propagation delays \cite[][]{1998ApJ...492L..49M}.  
Additionally, the pulsar shows evidence of intrinsic rotational irregularities (referred to as timing noise), which have previously made it difficult to identify secular variations in arrival times.

As a result, previous models have resulted in conflicting characterisation of the system.
\cite{1995ApJ...445L.137M} suggested that the propeller-driven torque close to periastron passage was causing arrival time variations.
\cite{1998MNRAS.298..997W} measured significant $\dot{x}$ and $\dot{\omega}$ and interpreted the results in the context of classical spin orbit coupling.  
\cite{2004MNRAS.351..599W} could not distinguish between discrete jumps in orbital parameters through periastron passage and long term variations associated with spin-orbit coupling.  For models including $\dot{x}$ and $\dot{\omega}$, \cite{2004MNRAS.351..599W}  found results highly inconsistent with   \cite{1998MNRAS.298..997W},  measuring values of $\dot{x}$ and $\dot{\omega}$ a factor of ten smaller in magnitude and opposite in sign. 

The system can also be constrained through studies of the companion star LS~2883. 
Recently, \cite{2011ApJ...732L..11N}   conducted optical spectroscopy of the companion. 
The companion mass was estimated to be $15$--$30 M_{\sun}$.  This higher mass implies a reduced orbital inclination angle of $ i \approx 23^\circ$ compared to previous studies (alternatively, the inclination angle could be $180-i \approx 157^\circ$) .   The broadening of absorption features in the spectra indicate that the companion is rotating at near-breakup velocity with an spin inclination of $\approx 30^\circ$ (or $150^\circ$) with respect to the line of sight.  
Observations of interstellar absorption lines together with  the high mass place the pulsar at a distance of $2.3 \pm 0.4$~kpc, much further than previous estimates of $\approx 1$~kpc.

\section{Observations}

The observations presented here were taken with the Parkes radio telescope between $1990$~January~18 and $2013$~February~3 and span six periastron passages of the system. 
 Early observations prior to $2003$ were taken with a series of analogue filterbank and digital autocorrelation spectrometers and are described in \cite{2004MNRAS.351..599W}.   
 Observations through the $2000$ periastron passage are presented in \cite{2002MNRAS.336.1201C}.
 Observations through the $2004$ periastron passage are presented in \cite{2005MNRAS.358.1069J}.  
  The cadence of these observations varies.   There are typically high-cadence observing campaigns close to periastron passage with observations many times per week.   At other times, monthly observing cadence is typically achieved.      
 Most recently, the pulsar has been observed with digital filterbanks as part of a programme to monitor pulsars of interest to the Fermi Space Telescope \cite[][]{2010PASA...27...64W}.   These observations have monthly cadence with a central observing frequency close to $1.4$~GHz and sem-iannual cadence with a central frequency of $3.0$~GHz.    
 
Close to periastron passage, the pulsar passes behind the dense wind of the companion and the pulsar shows a rapid increase in dispersion measure before being completely eclipsed by the star and its wind. 
We corrected the arrival times (TOAs) for variations in dispersion measure using published measurements of DM \cite[][]{2004MNRAS.351..599W,2005MNRAS.358.1069J}.  
We have corrected the TOA uncertainty for  uncertainties in the DM variation \cite[][]{cs2010}.
Arrival times were not corrected for the broadening of pulse profile associated with scattering of the pulsed radio emission in the equatorial disc and wind. 
For the most recent periastron passages we have estimated the dispersion measure by forming multiple TOAs across observed bandwidth.   

Formal  errors \cite[][]{1992RSPTA.341..117T} often underestimate the true TOA uncertainty because of a combination of astrophysical and instrumental effects.      
We find that the TOA uncertainties are underestimated for observations of this pulsar. 
Arrival time errors were therefore adjusted in order to better reflect white noise present in observations.  We did this by adding in quadrature an additional term to the formal TOA uncertainty.  For most of the observations, we have added $50~\mu$s to the TOA uncertainty, but  for some of the older observations, we have added $100~\mu$s because of the large observation-to-observation scatter in the measured arrival times.

\section{Timing Analysis}\label{sec:timing}

\subsection{Binary model}

The {\sc tempo2} software package \cite[][]{2006MNRAS.372.1549E} was used to model the pulse arrival times.  We used the planetary ephemeris DE421  and the most recently published realisation of terrestrial time (BIPM2011) to refer site arrival times to solar-system barycentre.    The binary orbit was modelled as a precessing Keplerian orbit using the  main sequence star (MSS) binary model \cite[][]{1998MNRAS.298...67W} initially implemented in {\sc tempo}  and re-implemented in {\sc tempo2}.
The  Keplerian orbital parameters are the epoch of periastron passage $T_0$, longitude of periastron passage $\omega$, eccentricity $e$, orbital period $P_b$, projected semimajor axis $x$.   
Precession of an orbit would cause slow variations in $x$ and $\omega$.   
We therefore extended the algorithm by fitting for  the first and second (time) derivatives of projected semimajor axis, respectively $\dot{x}$ and $\ddot{x}$, and  of longitude of periastron passage $\dot{\omega}$ and $\ddot{\omega}$.

There are other effects in the system  that can be manifested in TOA variations  \cite[][]{1998MNRAS.298..997W}.  We also considered models that included variation in the orbital eccentricity, $\dot{e}$,  and orbital period,   $\dot{P}_b$.
Finally we also considered the effects of Shapiro delay \cite[][]{1964PhRvL..13..789S}  and the orbital parallax effect \cite[][]{1995ApJ...439L...5K}.  This updated model is available in the {\sc tempo2} software repository.\footnote{\tt  tempo2.sourceforge.net}

\subsection{Time of arrival fitting algorithm}

Our fitting method differs from previously applied methods in a few important ways. 
Firstly, we have incorporated a red noise covariance matrix into the TOA fitting process using a method described in  \cite{2011MNRAS.418..561C}.
The red noise was estimated using an iterative procedure. From a first estimate of the binary orbital properties, we calculated the post-fit residuals.
These residuals show evidence for red noise that cause variations in the pulse arrival time that vary over $\approx 20$~cycles of pulse phase over the $23$~yr of observation.  
We then estimated the power spectrum of these residuals, as displayed in Figure \ref{fig:spectrum}. 
The power spectrum was calculated using a pre-whitening technique to mitigate the effects of spectral leakage of red-noise processes. 
We found that the spin noise could be modelled as a red noise process with power spectral density $P(f) \propto f^{-8}$. 
This level of timing noise is consistent with that of other rotation-powered pulsars \cite[][]{sc2010}.
The red noise was incorporated into the covariance matrix when the model was fitted. 
The fitting process and noise modelling process were iterated until the fitted parameters and the noise model converged.

The best-fitting residuals, displayed in Figure \ref{fig:residuals}, show that the timing solution is reasonable because there are no pulse-number ambiguities at any time, meaning the solution is phase-coherent.  
Additionally, the variations appear to be smooth,  indicative of a red noise process, and lacking quasi-periodic variations at the orbital period of the system.
 The plot does not provide visual or quantitative tools necessary to assess the goodness of fit.
It is difficult to visually detect the sub-millisecond variations caused by spin-orbit coupling. 
Finally, the root-mean-square level of the residuals, which is often used to measure assess the goodness of fit, is not an appropriate diagnostic here. 

In order to both visually and quantitatively assess the check the goodness of fit we analyse the whitened residuals,
 \be
\bm{w} =   \textbfss{u}^{-1} \bm{r},
 \ee
 where the whitening matrix $\textbfss{u}^{-1}$ is calculated from the Cholesky decomposition  of the noise covariance matrix $\textbfss{C}$ (i.e, $\textbfss{u}  \textbfss{u}^t \equiv \textbfss{C}$) and $\bm{r}$ is a column vector of the best-fitting residuals. 
 
If the covariance matrix adequately models the  timing noise and the timing model accounts for all of the deterministic TOA variations, the whitened residuals should show white-noise characteristics and have an rms of unity \cite[][]{2011MNRAS.418..561C}.
   The quality of the fits can be assessed using the goodness of fit statistic
   \be
\chi^2 = \bm{r}^T \textbfss{C}^{-1} \bm{r}.
 \ee 
If both the timing and noise models are good, the reduced-$\chi^2$  $\chi^2_R \equiv \chi^2/\rm N_{\rm DOF} \approx 1$, where $N_{\rm DOF}$ is the number of degrees of freedom in the fit (i.e., the difference between the number of data points and the number of fit parameters).  

We considered different models for the TOAs that contained different numbers of parameters. 
Including additional parameters in the fit improves the model and will reduce the  $\chi^2$ value, even if the parameters are not significant.  
To test if the reduction in $\chi^2$ associated with the inclusion of a given parameter was significant, we used the Akaike Information Criterion \cite[][]{1100705} to determine if the improvement was significant, requiring that  the $\chi^2$ in the new fit to be smaller than the original $\chi^2$ by a value of  $2$ for each parameter added to the fit. 

\subsection{Results of fitting}

We compared three different families of models:  1) Keplerian parameters only;  2) the addition of post-Keplerian parameters; and  3) Keplerian, post-Keplerian, and proper motion. 
 
In Table \ref{tab:model_comp}, we show the $\chi^2$ and $\chi^2_R$  values for the models we considered.    
 We find that including the first and second derivatives of $x$ and $\omega$ provides a significantly improved fit with the reduced $\chi^2$ decreasing from $7$ to $2.5$ and the formal uncertainties on $\dot{x}$ and $\dot{\omega}$ implying, respectively, $25\sigma$ and $40\sigma$ detections of these parameters.
 These improvements can also be found by comparing the power spectra of the solutions.
We also compare solutions that have had one extra cycle of pulse phase added and one extra cycle of pulse phase subtracted across each periastron passages.  These solutions show poorer $\chi^2$ than the best-fiting solution.

The best-fitting model is shown in Table \ref{tab:params}.
 Our measurements of $\dot{x}$ and $\dot{\omega}$ and their uncertainties are an order of magnitude smaller than those previously published.
 Also, for the first time, we have detected significant values for  $\ddot{\omega}$ and $\ddot{x}$ with the detections of both parameters showing $ > 3\sigma$ statistical significance.   
We also measured significant orbital period increase $\dot{P_b}$, and proper motion in both right ascension $\mu_\alpha \cos \delta$ and declination $\mu_\delta$.     
The implications of these measurements are discussed in Section \ref{sec:implications}.

In Figure \ref{fig:whitened_residuals}, we show the residuals  whitened using our model for the timing noise for a model that contains only the Keplerian parameters (panel {\em a}) and the best-fitting model (panel {\em b}).
For all of the models considered,  we found that there is a small ($\ll 0.005$ cycle of pulse phase) systematic 
 variation in the residual arrival times close to periastron passages.  These departures are biased to positive
  contribution to arrival time. They could therefore be  plausibly associated with uncorrected variations in dispersion measure or scattering of the pulsar radiation through the disc \cite[][]{1998ApJ...492L..49M}. 
   Close to periastron passage, they find that the scintillation bandwidth at $4.8$~GHz is approximately $\nu_{d, 4.8} \approx 5$~MHz.    Assuming Kolmogorov turbulence, the scintillation bandwidth scales $\propto \nu^{-4.4}$ and the scintillation bandwidth at $1.4$~GHz is  $22$~kHz. The scatter-broadening time can be computed from the scintillation bandwidth using $\tau_d \approx 2 \pi/\nu_d$  \cite[][]{1999ApJ...517..299L}.  
   Therefore at $1.4$~GHz we expect $\tau_d \approx 2 \pi/\nu_d \approx 0.28~$ms     . 
In the regime relevant here where scattering is small compared the pulse width the delay in the arrival time is the scatter broadening time \cite[][]{cs2010}.      We find that scattering delays induce delays of approximately $0.006$ cycles of pulse phase ($0.29$~ms) close to eclipse.

\begin{figure}
\begin{center} 
\begin{tabular}{c}
\includegraphics[scale=0.5]{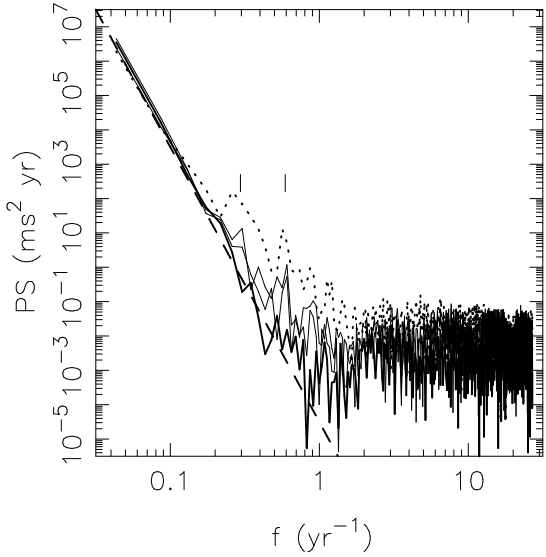}  \\
\end{tabular}
\caption{  \label{fig:spectrum}  \footnotesize Power spectrum  for observations of PSR~B1259$-$63 (solid thick line).  The dotted line shows the power spectrum derived from a model that only includes Keplerian parameters.  The thin solid lines show models that include additional phase jumps across periastron passages.   These models show evidence for excess power, particularly at the fundamental and first harmonic of the orbital period, which are highlighted by vertical solid lines. 
The dashed line shows the $f^{-8}$ model for the noise, used both as a prewhitening filter to estimate pulsar parameters and the basis for simulations to test the robustness of parameter estimation.    }  \end{center} 
\end{figure}

\begin{figure}
\begin{center} 
\begin{tabular}{c}
\includegraphics[scale=0.5]{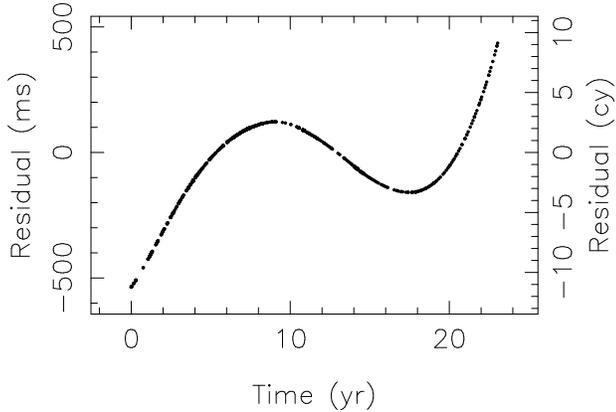}  \\
\end{tabular}
\caption{  \label{fig:residuals} Residual times of arrival for the best-fitting model.  }  \end{center} 
\end{figure}

\begin{figure}
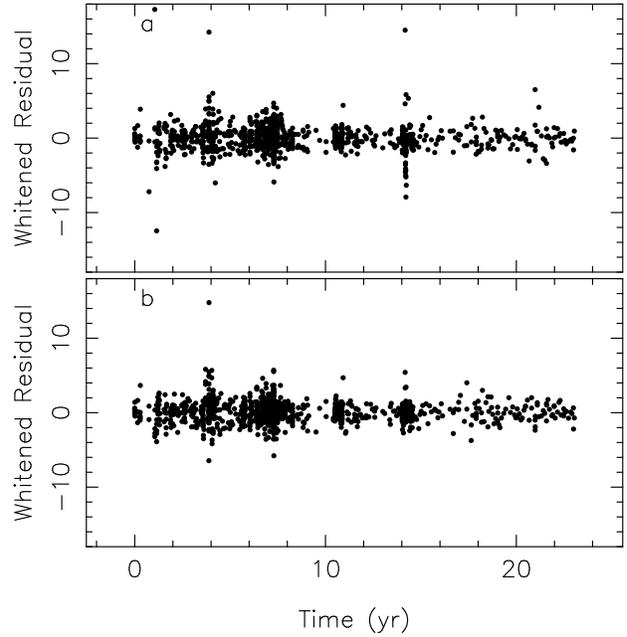

\begin{center} 

\includegraphics[scale=0.6]{keplerian.eps}  \\
\includegraphics[scale=0.6]{all_white.eps}  
\caption{ \footnotesize \label{fig:whitened_residuals} Whitened residuals for timing models of PSR~B1259$-$63. Panel a):  Only Keplerian parameters; b) The best-fitting model.  The whitened residuals are dimensionless (and hence the ordinate of the plot) are dimensionless.   }  \end{center}
\end{figure}

In Figure \ref{fig:spectrum}, we show the power spectrum for the best-fitting Keplerian solution,   which has excess power  associated with improperly modelled TOAs.
 We also show the best solutions with one extra cycle of pulse phase added and one extra pulse phase subtracted at each periastron passage.  The excess in the $\chi^2$ can be attributed to excess power at the fundamental and first harmonic of the orbital period.

For other parameters, we did not measure significant values.
In particular, \cite{1995ApJ...439L...5K} suggested that with $\approx 100~\mu$s timing precision, orbital parallax may be detectable for this pulsar because of the wide orbit (making this effect large).   Given the distance to the pulsar we expect $\pi_{\rm orb} = 1~$mas.   When this term is include in the fit, we found an insignificant improvement and set an upper limit of $\pi_{\rm orb} < 6$~mas, well above the expected value.

This pulsar also experienced a small glitch close to the third periastron passage that has previously been identified and characterised \cite[][]{2004MNRAS.351..599W}.   
We have re-fit for all of the glitch components.  Our value for permanent frequency change is consistent with that published in \cite{2004MNRAS.351..599W}.  We find no evidence for a permanent change in the frequency derivative.    Because we have fitted for the glitch components in the presence of the red-noise timing model, our glitch parameters are likely more robust than previous estimates.

\subsection{Estimating parameter uncertainty and ruling out parameter bias} \label{sec:param_uncertainty_bias}

The best model has a reduced $\chi^2$ value of approximately $2$.  This indicates that the noise model does not fully account for the TOA uncertainty.  We attribute the excess to systematic variations close to periastron passage and unmodelled instrumental effects.     In order to determine if there was bias in our parameter estimation and if  parameter uncertainties were being underestimated, we conducted two sets of simulations.

In one set of simulations, we assessed if parameter estimation was being affected by the red noise and red-noise modelling algorithm we employed.    
To do this, we generated synthetic datasets that matched the noise model for the observations.
The data sets were formed by adding to ideal TOAs that perfectly matched the pulsar ephemeris,  red and white noise consistent with the observations.   We then re-fit the pulsar spin model to these datasets.  
We found that our results were independent of starting from the true value or starting from values significantly different than the true value (for example, by starting the modelling process with parameters of interest, such as  $\dot{x}$, set to zero).  
 
We measured the variance of the measured values from each simulation as well as the reduced $\chi^2$ of the ensemble:   
\be
\label{eqn:chi_red}
\frac{\chi^2}{N_{\rm DOF}} = \frac{1}{N_{\rm sim}-1} \sum_{i=1}^{N_{\rm sim}} \left(\frac{x_i - \langle x \rangle} {\delta x_i} \right)
\ee 
where $\langle x \rangle$ is the average of the $N_{\rm sim}$ realisations, $x_i$ are the values from individual realisations, and $\delta x_i$ are the inferred parameter errors.   
If the parameter errors are correct, we would expect $\chi^2/N_{\rm DOF} \approx 1$.  In Table \ref{tab:chi_cor} we show $\chi^2$ values. We found that for all of the parameters, there was no bias in the parameter estimation.  
We found that in all cases, the  reduced $\chi^2 / N_{\rm DOF} < 1.9$, indicating that the error bars for all parameters were correct to within a factor of $ \approx  \sqrt{\chi^2} < 1.4$.  

To assess the presence of statistical bias in our methods, we used  the statistic $\Delta / \sigma  \equiv (p - \langle p \rangle)/\sigma_p $, where $\sigma_p$ is the standard deviation of the parameter $p$, and $\Delta = p -\langle p \rangle$ is the difference bewteen the actual parameter value and the mean of the $N_{\rm sim}$ simulations. 
We found that for all parameters, the bias parameter is consistent with the predicted value of  $| \Delta /\sigma| \lesssim 1/\sqrt{N_{\rm sim}} = 0.1$.  

In a second set of simulations, we used a bootstrap method to assess if the under-estimation of the white noise levels is biasing the measurement of parameters or their uncertainties.
  We formed $100$ realisations by randomly selecting TOAs (with replacement) from the original list of TOAs. 
We then re-fit the timing model to each realisation.
  We then calculated the reduced-$\chi^2$  values (Equation \ref{eqn:chi_red}), which are also presented in   Table \ref{tab:chi_cor}. 
We find again that for all parameters the reduced $\chi^2 < 2$.  In these simulations the bias parameter has slightly different meaning than in the red noise simulations.   Because the input value is not the true value, but only a value derived from a single realisation, we expect $\Delta /\sigma_p \approx 1$ for all of the parameters.  This is consistent with the estimates for the bias parameter that we derive.

From these simulations we conclude that our parameter estimation is not biased and the uncertainties  in  all cases are correct to within a factor of $\approx 1.4$ at worst. 
 In subsequent analysis, we correct the parameter uncertainties  by multiplying them by $\sqrt{\chi_R^2/{N_{\rm DOF}} + \chi_W^2/{N_{\rm DOF}}}$, using  values listed in Table \ref{tab:chi_cor}. 
 

\begin{table}
 \centering
 \caption{Model Comparison.  $K$ refers to a fit to the Keplerian parameters.    \label{tab:model_comp}}
\begin{tabular}{lcc}
\hline \hline
Model & $\chi^2$  & $\chi^2/N_{\rm DOF}$ \\ \hline

$K$ &  $3363$  &    $2.70$ \\
$K$ + $\dot{x}$ + $\dot{\omega}$  &  $2515$  &   $2.02$ \\
$K$ + $\dot{x}$ + $\dot{\omega}$  + $\mu$  &   $2420$  &   $1.94$\\
$K$ + $\dot{x}$ + $\dot{\omega}$  + $\mu$+ $\ddot{x}$ + $\ddot{\omega}$ &   $2416$  &   $1.94$\\
$K$ + $\dot{x}$ + $\dot{\omega}$  + $\mu$ +  $\ddot{x}$ + $\ddot{\omega}$ + $\dot{P}_b$    & $2382$  &   $1.91$\\
\hline
\end{tabular}
 
\end{table}

\begin{table}
\begin{center}
 \caption{ Results of simulations.    \label{tab:chi_cor}  The columns labelled {\em R}  show the results from the red-noise simulations.  The columns labelled {\em W} show the reduced results from the bootstrap methods used to estimate the effects of underestimation white noise.  } 
\begin{tabular}{crrrr}
\hline \hline
& \multicolumn{2}{c}{R} & \multicolumn{2}{c}{W}  \\ 
 Parameter & $\Delta /\sigma $& $\chi_R^2/N_{\rm DOF}$ & $\Delta /\sigma $& $\chi_W^2/N_{\rm DOF}$ \\ \hline
$\dot{x}$ & 		 $-$0.02 & 1.1 & $-$0.5 & 1.2  \\
$\ddot{x}$ & 		 0.02 & 1.0 &  $-$0.3 & 1.0 \\
$\dot{\omega}$ & 0.06 & 1.4 &  $-$0.3 & 1.3 \\
$\ddot{\omega}$& 0.1 & 1.2 &   $-$0.002 &  1.3\\
$\dot{P}_b$&   0.08 & 1.3 &  $-$0.4  & 1.7 \\
$\mu_\alpha \cos \delta$ &  $-$0.06 & 1.9 & $-$0.5 & 0.8\\
$\mu_\delta $& $-$0.09&  1.4 &$-$0.3  &0.9 \\
\hline
\end{tabular}
\end{center}
 \end{table}

\begin{table*}
\begin{centering}
\caption{Best-fitting parameters for PSR~B1259$-$63. The spin frequency and frequency derivative parameters presented here were fit while the other parameters were held fixed without use of the Cholesky prewhitening algorithm.  $\nu$ and $\dot{\nu}$ were held fixed when the other parameters were fit. }
\label{tab:params}
\begin{tabular}{ll}
\hline
\hline
\multicolumn{2}{c}{Measured Quantitites} \\ 
\hline
Right ascension, $\alpha$\dotfill &  13:02:47.6426(11) \\ 
Declination, $\delta$\dotfill & $-$63:50:08.665(8) \\ 
Pulse frequency, $\nu$ (s$^{-1}$)\dotfill & 20.93692420(15) \\ 
First derivative of pulse frequency, $\dot{\nu}$ (s$^{-2}$)\dotfill & $-$9.989(5)$\times 10^{-13}$ \\ 
Proper motion in right ascension, $\mu_{\alpha} \cos \delta$ (mas\,yr$^{-1}$)\dotfill & $-$6.6(1.8) \\ 
Proper motion in declination, $\mu_{\delta}$ (mas\,yr$^{-1}$)\dotfill & $-$4.4(1.4) \\ 
Orbital period, $P_b$ (d)\dotfill & 1236.724526(6) \\ 
First derivative of orbital period, $\dot{P_b}$\dotfill & 1.4(7)$\times 10^{-8}$ \\ 
Epoch of periastron, $T_0$ (MJD)\dotfill & 53071.2447290(7) \\ 
Projected semi-major axis of orbit, $x$ (lt-s)\dotfill & 1296.27448(14) \\ 
First derivative of $x$, $\dot{x}$  (lt-s s$^{-1}$)\dotfill & 2.100(8)$\times 10^{-11}$ \\ 
Second derivative of $x$, $\ddot{x}$ (lt-s s$^{-2}$) \dotfill & 2.0(7)$\times 10^{-20}$ \\ 
Longitude of periastron, $\omega_0$ (deg)\dotfill & 138.665013(11) \\ 
Orbital eccentricity, $e$\dotfill & 0.86987970(6) \\ 
Periastron advance, $\dot{\omega}$ (deg~yr$^{-1}$)\dotfill & 7.81(3)$\times 10^{-5}$ \\ 
Periastron acceleration $\ddot{\omega}$ (deg~yr$^{-2}$)\dotfill & 1.8(3)$\times 10^{-6}$ \\ 
Glitch epoch \dotfill & 50708.0(5) \\ 
Glitch frequency permanent frequency change \dotfill & 3.1(3)$\times 10^{-8}$ \\ 
Glitch transient frequency change \dotfill & 1.8(3)$\times 10^{-8}$ \\ 
Glitch transient decay time\dotfill & 82(18) \\ 
\hline
\multicolumn{2}{c}{Fixed Quantities} \\ 
\hline
Epoch of frequency determination (MJD)\dotfill & 50357 \\ 
Epoch of position determination (MJD)\dotfill & 55000 \\ 
Epoch of dispersion measure determination (MJD)\dotfill & 50357 \\ 
Dispersion measure, DM (pc~cm$^{-3}$)\dotfill & 146.8 \\ 
\hline
\multicolumn{2}{c}{Derived Quantities} \\
\hline
$\log_{10}$(Characteristic age, yr) \dotfill & 5.52 \\
$\log_{10}$(Surface magnetic field strength, G) \dotfill & 11.52 \\
\hline
\multicolumn{2}{c}{Assumptions} \\
\hline
Clock correction procedure\dotfill & TT(BIPM2011) \\
Solar system ephemeris model\dotfill & DE421 \\
\hline
\end{tabular}
\end{centering}
\end{table*}

\section{Implications} \label{sec:implications}

\subsection{Spin-orbit coupling}

Our best-fitting model includes significant measurements of the first and second derivatives of the parameters $x$ and $\omega$.  We attribute these measurements to spin-orbit coupling.    
These measurements can therefore be used to determine the spin-orbit misalignment and place strong constraints on  system geometry and evolution.

 \cite{1998MNRAS.298...67W} derives expressions for the first and second derivatives of $x$ and $\omega$ in the presence of spin-orbit coupling:
\be
&&\dot{x} = n Q x \cot i \sin \theta \cos \theta \sin \Phi_0  \label{eqn:xdot}\\
&&\ddot{x} = - n^2 Q^2 x \cot i \left( \frac{\sin^2 \theta \cos^2 \theta \cos \Phi_0}{\sin \theta_J}\right) \label{eqn:x2dot} \\
&&\dot{\omega} = n Q \left( 1 - \frac{3}{2} \sin^2 \theta + \cot i \sin \theta \cos \theta \cos \Phi_0\right) \label{eqn:omdot} \\\
&&\ddot{\omega} = n^2 Q^2 \cot i \left( \frac{\sin^2 \theta \cos^2 \theta  \sin \Phi_0}{\sin \theta_J} \right) \label{eqn:om2dot}.
\ee
The factor $Q$ parametrizes the quadrupole moment of the companion
\be
\label{eqn:mom_init}
&&Q = \frac{3 J_2/M_c}{2a^2(1-e^2)^2} =  k \frac{ R_c^2 \hat{\Omega}_c^2 }{a^2(1-e^2)^2} 
\ee
where  $J_2 = I_3- I_1$,  $I_3$ is the moment of inertia about the symmetry axis,  $I_1$ is the moment of inertia about an equatorial axis, $k$ is the apsidal-motion constant and $ \hat{\Omega}_c \equiv \Omega_c/(G M_c/R_c^3)^{1/2}$
  is the dimensionless proper rotation of the companion, $a$ is the semi-major axis of the relative orbit, and $e$ is the orbital eccentricity.  
We note that it is necessary to correct $\dot{\omega}$ for a relativistic contribution that  is discussed in Section \ref{sec:implications_gr}

\begin{figure}
\begin{center} 
\includegraphics[scale=0.7]{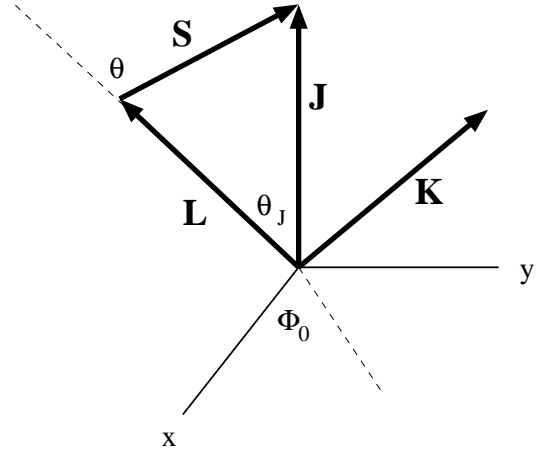}  \\
\caption{ \footnotesize \label{fig:orbit_geometry} Orbital Configuration.  The total angular momentum $\bm{J} = \bm{L} + \bm{S}$  is the sum of the orbital angular momentum $L$ and the spin angular momentum $\bm{S}$.  $\bm{J}$ is defined to be in the $\bm{\hat{z}}$ direction and the observer's line of sight is defined to be in the $x$-$z$ plane. Not shown on the figure are, $i_L$, and $i_S$, respectively the inclination angles of the orbital angular momentum  $\bm{L}$  and the spin angular momentum $\bm{S}$ with respect to the line of sight. }  \end{center}
\end{figure}

The spin-orbit geometry can be characterised by four angles: $\Phi_0$, $\theta$, $\theta_J$,  which are shown in Figure \ref{fig:orbit_geometry}, and the inclination angle $i$ of the total angular momentum $\bm{J} = \bm{L} + \bm{S}$, where $\bm{L}$ is the orbital angular momentum and $\bm{S}$ is the spin angular momentum.    
Following \cite{1998MNRAS.298...67W}, we define a coordinate system such that the total angular momentum $\bm{J}$ is in the $\hat{z}$ direction and the observing line of sight is in the $y$-$z$ plane, inclined from $\hat{z}$ by an angle $i$.     
The angle between $\bm{L}$ and $\bm{J}$ is $\theta_J$, and the angle between $\bm{L}$ and $\bm{S}$ is $\theta$.
The ascending node of the orbital angular momentum on the total angular momentum plane at the epoch of the measurements of $\dot{x}$ and $\dot{\omega}$ is  $\Phi_0$. 
Because $\bm{S} \ll \bm{L}$, $\theta_J \ll 1$ and $i$  is to good approximation  the inclination of the orbit with the line of sight $i_L$.  The inclination of the spin of the companion is $i_S$.

\begin{figure}
\begin{center} 
\includegraphics[scale=0.5]{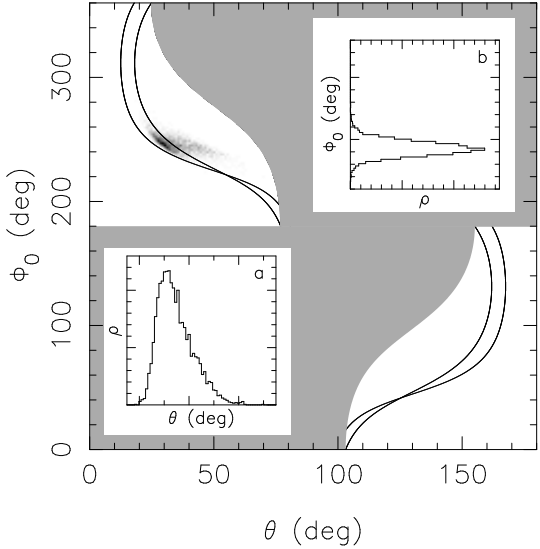}  \\
\caption{ \footnotesize \label{fig:theta_phi0} Main panel:  Allowed values of $\Phi_0$ and $\theta$.   The lines define the region allowed based on measurements of $\dot{x}$ and $\dot{\omega}$.  The allowed values have a range because of uncertainties in the inclination of the system.   The solid grey region is excluded because $\dot{x} > 0$ and $\dot{\omega} > 0$.   The grey shaded region shows additional constraints that come from measurements of $\ddot{x}$ and $\ddot{\omega}$.  Insets:  distributions for $\theta$ (panel a) and $\Phi_0$ (panel b), based on the best-fitting solution.   In the insets, the ranges of $\theta$ and $\Phi_0$ are drawn to scale with the main panel.         }  \end{center}
\end{figure}

We first examine constraints that can be placed on the orbital parameters using measurements of $\dot{x}$ and $\dot{\omega}$.  We do this both because these two quantities are measured with the highest significance and they have previously been analysed, and therefore enable direct comparison of our results to previous work \cite[][]{1998MNRAS.298..997W,2004MNRAS.351..599W}.
The periastron advance $\dot{\omega}$ needs to be corrected for the general relativistic contribution; this correction  is discussed in Section \ref{sec:implications_gr}.
 Using these measurements of $\dot{x}$ and $\dot{\omega}$, we can solve Equations~(\ref{eqn:xdot})~and~(\ref{eqn:omdot}) to constrain values of $\Phi_0$ and $\theta$:
\be
\label{eqn:constraint_theta_phi}
\frac{\dot{\omega}  x}{\dot x} \sin \Phi_0  - \cos \Phi_0 = \frac{1 + 3 \cos 2\theta}{2 \cot i \sin 2\theta}.
\ee 
In Figure \ref{fig:theta_phi0} these constraints are shown as the solid black curves. 
The widths of the band correspond to the uncertainty in the inclination angle $i$  of  $10^\circ$.    
  The solid grey region is excluded because  $\dot{x} > 0$  and $\dot{\omega} > 0$.   Based on these measurements alone, we find that the inclination of the spin angular momentum with respect the orbital angular momentum has the range $   15 ^\circ \lesssim \theta \lesssim 165 ^\circ $.        
There are two sets of allowed values of $\theta$, as Equation (\ref{eqn:constraint_theta_phi}) enables values of $\theta$ that represent approximately parallel and  anti-parallel $\bm{L}$ and $\bm{S}$.
However, from expectations for the evolution of binary star systems we expect $\bm{L}$ and $\bm{S}$ to be closer to parallel than anti-parallel so we favour the permissible region in the first quadrant of Figure \ref{fig:theta_phi0}.

The additional measurements of $\ddot{x}$ and $\ddot{\omega}$ allow  the three additional angles $\Phi_0$, $\theta$, and $\theta_J$ to be determined.
With measurements of these values we are also able to determine the inclination of the spin-axis with respect to the line of sight. 

Monte-Carlo simulations were used to estimate the uncertainty for the parameters. 
We simulated values of $\dot{x}$, $\ddot{x}$, $\dot{\omega}$, and $\ddot{\omega}$ from  distributions consistent with their measured uncertainties. In these simulations, we also incorporated the covariance between the parameters as determined from the fit.  

We simulated values of the companion mass  from the estimates of \cite{2011ApJ...732L..11N}.  We fix the pulsar mass at $1.4 M_{\sun}$.  This is justified because the uncertainties in neutron star masses (particularly those of non-recycled pulsars like PSR~B1259$-$63) are much smaller than those of the companion, and the kinematics  are dominantly affected by the large mass of the companion and mass of the progenitor.      

We then solved Equations \ref{eqn:xdot} to \ref{eqn:om2dot} to calculate $\theta$, $\theta_J$, $\Phi_0$ and $Q$.  The derived values for these quantities are presented in Table \ref{tab:geo_param}, with the best matching values of $\Phi_0$ and $\theta$ presented as the shaded grey region in Figure \ref{fig:theta_phi0}.
   As expected, we find that the vast majority of total angular momentum is contained in the orbital angular momentum; therefore $\theta_J \ll 1$ and $\bm{L}$ is essentially parallel to $\bm{J}$.  We also find that the spin of the companion star is misaligned from total angular momentum by $\approx 35^\circ$. 

By combining our results with optical spectroscopy of the companion, we can calculate the apsidal-motion constant for the star. 
Optical spectroscopy supports a fast-rotating companion with the projected velocity $v_{\rm rot} \sin i_S = 260 \pm 15$ km~s$^{-1}$  and a radius of $\approx 10~R_{\sun}$ \cite[][]{2011ApJ...732L..11N}.   Our estimate of the spin inclination $i_S \approx 147^\circ$  is in good agreement with values derived from optical observations \cite[][]{2011ApJ...732L..11N}.
Combining radio and optical measurements,  we find that $\hat{\Omega} \approx 0.8$, indicating that the star is spinning at near breakup velocity. 
We can solve Equation \ref{eqn:mom_init} to find the apsidal-motion constant is $ k \approx 10^{-3}$.  
This is consistent with the low surface gravity of the companion \cite[$\log g \approx 4$,][]{1999A&A...350...56C}.  
There is large $0.8$~dex uncertainty in $k$  because of the compounded  uncertainties in companion mass and rotational speed. 
We find that  the difference between the on-axis and equatorial moments of inertia is $J_2 \approx 5\times10^{54}$~g~cm$^{2}$.

\begin{table}
\centering
\caption{Orbital parameters derived from spin-orbit coupling\label{tab:geo_param}  }
\begin{tabular}{ccc}
\hline \hline
Parameter  & Value & Definition  \\ \hline
$\Phi_0$  & $245(6)^\circ$  & Figure \ref{fig:orbit_geometry} \\
 $\theta$  & $ 35(7)^\circ$  & Figure \ref{fig:orbit_geometry} \\
$\theta_J$ & $ 3(1)^\circ\times 10^{-4}$  & Figure \ref{fig:orbit_geometry} \\
 $Q$ &    $3.5(7)\times 10^{-7}$  & Equation \ref{eqn:mom_init}\\ \hline
$J_2/M_c$  & $5.2(3) \times 10^{-7}$~au$^2$ & Equation \ref{eqn:mom_init} \\
   $i_L$ & $153(4)^\circ$  & Figure 4 \\
 $i_S$ &  $147(3)^\circ$ &  Figure 4 \\ \hline
 $\log k$ & $-3.0  \pm0.8 $ & Equation \ref{eqn:mom_init}  \\ \hline 
 \end{tabular}

\end{table}

\subsubsection{Comparison with eclipse geometry}

The eclipse of the pulsed pulsar radiation is associated with the passage of the star behind the equatorial excretion disc.   
\cite{1995MNRAS.275..381M} found that the eclipse will only occur if orbital plane is offset from the equatorial plane of the companion star by an angle of between $10^\circ$ and $40^\circ$.    Our measurement of an inclination angle of $\theta \approx 35^\circ$ is consistent with these values.

\subsubsection{Emission  geometry:  evidence for spin-spin alignment}

The geometry of the pulsar magnetosphere is consistent with the spin of the neutron star progenitor being aligned with the companion prior to the supernova explosion.

  \cite{1995ApJ...441L..65M} present an analysis of the polarisation of the pulsar.  
 They interpreted the two components of the pulse profile as being associated with components of a wide beam.   
 The position angle sweep was interpreted within the context of the rotating-vector model \cite[][]{1969ApL.....3..225R} allowing the measurement of the angle between the magnetic axis and rotation axis $\alpha$, and the angle between the rotation axis and the line of sight $\zeta$.    They presented two possible models for the emission geometries\footnote{We have converted the models to represent the convention for emission angles presented in \cite{2001ApJ...553..341E} that has since been adopted.}:  one with ($\zeta = 170^\circ $, $\alpha = 170^\circ$)  (no uncertainties were presented for this fit in \citealt{1995ApJ...441L..65M})  and a second with ($\zeta=134^\circ \pm 6^\circ$, $\alpha = 137^\circ \pm 0.3^\circ$). The latter interpretation was favoured because it better fitted the data and  the  pulse profile evolution with frequency.  We have also revisited the polarimetry using more recent high time and frequency resolution observations and have confirmed the \cite{1995ApJ...441L..65M} results. 
 
The timing measurements suggest that the companion spin  has an inclination with respect to the line of sight of $147^\circ \pm 3^\circ$.
Polarimetry of the pulsar suggests that the spin angular momentum of the pulsar has a similar inclination with respect to the line of sight.  However, the position angle  on the sky is not constrained.  


\subsection{Proper motion: evidence of association with Centaurus OB1}

The significant proper motion can be used to identify potential birth location for the system and, if the location is identified, make an estimate of the system's kinematic age.   
We find that the pulsar has a proper motion of $(\mu_\alpha \cos \delta, \mu_\delta) = (  -6 \pm  2  ,  -4 \pm 1)$~mas~yr$^{-1}$. 
\cite{2011ApJ...732L..11N} identified interstellar absorption lines in the spectrum of the companion to estimate  the distance to be $2.3 \pm 0.4$~kpc.   Using this assumed distance, we find a transverse velocity of  $\approx 80\pm30$~km~s$^{-1}$.
 
\cite{2011ApJ...732L..11N} also noted that the system has extinction consistent with members of the Cen OB1 association.
The centre of the association is $(\ell,b) = (303^\circ.7, 0^\circ.5)$, which corresponds to $(\alpha, \delta) = (12^{\rm h}58^{\rm m}, -62^\circ28^{'})$ with an angular extent of $4 \times 4 \deg^2$. 
In $3\times 10^5$~yr, the pulsar has moved $1.1^\circ$ in right ascension and $0.4^\circ$ in declination.
Correcting for systemic motion of the cluster, the  birth position of the pulsar is $(\ell, b) = (303^\circ.9, -0^\circ.6)$, closer to the centre of the association than its current location.  The kinematic age of the system, assuming the system originated close to the centre of the cluster, is therefore consistent with the spin-down age.

A large portion of the system proper motion can be attributed to the cluster. 
\cite{2013A&A...553A.108C} note that the Centaurus OB association has a proper motion of $\mu_\alpha \cos \delta = -4.8 \pm 0.1$~mas~yr$^{-1}$ and $\mu_\delta =0.8 \pm 0.1$~mas~yr$^{-1}$ . 
Correcting for the cluster motion, the system has a speed of  $60 \pm 30 $~km~s$^{-1}$, which is consistent with a population of OB runaway stars \cite[][]{2011MNRAS.410..190T}
We also note that the trajectory of the star is inconsistent with the star originating in the young open cluster NGC~1755.

There are inconsistent measurements of the optical radial velocity of the companion star. 
\cite{1994MNRAS.268..430J} measured a blue-shift of H$\alpha$ emission of the companion that of $70$~km~s$^{-1}$ relative the local standard of rest.   
However \cite{2011ApJ...732L..11N} do not measure a significant shift of the H$\alpha$ emission relative to the local standard of rest.   We note that orbital radial velocity variations cannot explain this discrepancy.   The star shows approximately $\approx 10$~km~s$^{-1}$ peak-to-peak radial velocity variations \cite[][]{1994MNRAS.268..430J}, which occur predominantly near periastron passage.   Because both observations occurred far from periastron passage, the  orbital effects are much smaller than $10$~km~s$^{-1}$.  In the following section, we consider both radial velocities in modelling of the system.

\subsection{Three-dimensional orbit of the system: constraining progenitor masses and supernova kicks}

The model of the current orbit of the system can be used to constrain the properties of the system at the time of the supernova explosion.

We let the pulsar progenitor have an unknown initial mass $M_1$ and an orbit with semimajor axis $a_0$.   The orbit is assumed to be circular and in the plane perpendicular to the spin of the companion.  
These assumptions are reasonable because the orbit of the stars is expected to be circularised by gas friction prior to the supernova explosion. 
The spin of the companion is likely aligned with the orbit because of mass transfer prior to the supernova explosion \cite[][]{1984JApA....5..209V,1995MNRAS.274..461B}.  We also assume that prior to the supernova explosion, the system was moving with a velocity equal to that of the average of the Cen OB1 cluster.  
 
 In the supernova explosion, the progenitor to the neutron star loses $M_1-M_{\rm NS}$ of its mass, where $M_{\rm NS} = 1.4 M_{\sun}$ is assumed for the pulsar mass.    Immediately after the supernova explosion, the companion velocity is its orbital velocity prior to the explosion.  The pulsar's velocity is the sum of its circular velocity and a kick velocity $\bm{V}_{\rm kick}$. 
 This kick is assumed to be instantaneous (i.e., it is imparted on a timescale much shorter than the post-explosion orbital time scale).  This kick could arise from different sources: it could be either associated with an asymmetric explosion, or acceleration of the neutron star with a variety of possible mechanisms  \cite[][]{2004cetd.conf..276L,2006ApJ...639.1007W}. 

The input parameters for the model are displayed in Table \ref{tab:constraints}.   The orbit of the post-supernova system and its centre of mass velocity can be   characterised by the masses, initial positions, and initial velocities of pulsar and its companion. 
Expressions for these quantities are derived in Equations (2.134) to (2.139)  of \cite{1999ssd..book.....M}.
 The three-dimensional space velocity of the system is the centre-of-mass velocity. 
The orbit of system is characterised by its semimajor axis, inclination, eccentricity, longitude of ascending node, longitude of periastron, and mean anomaly at the time of the supernova explosion. 
These parameters represent the orbital configuration at the time of the explosion;  the parameters have certainly evolved since the explosion (e.g., we know that the orbit is precessing due to spin-orbit coupling).

Given the weak or unconstrained nature of many of the input parameters, it is necessary to search over a large parameter space in order to find values compatible with the observed system.
We used Monte-Carlo methods to search the parameter space. 
To do this, we first randomly generated input parameters from broad distributions, consistent with observational and astrophysical constraints. 
Using these initial parameters, we then calculated the post-explosion properties of these simulated systems and compared them to the observed system.
The match was assessed using the goodness-of fit criterion
\be
-\log L   =  \sum_{i=1}^{N_p}  \left(\frac{p_{i,{\rm sim}} -p_{i,{\rm obs}}  }{\delta p_{i,{\rm obs}}} \right)^2
\ee 
where $\bm{p}=(p_1, ... p_{N_p})$ is a vector containing the parameters used to constrain the system and $\bm{\delta p}$ are their uncertainties.  
A good match is found when  when $ -\log L  \approx 1$. 

For constraining parameters, displayed in Table \ref{tab:constraints}, we used the current three-dimensional space velocity $\bm{v}$, the projected semimajor axis of the pulsar's orbit $x$, the orbital period $P_b$, the orbital eccentricity $e$, and the angle between the spin and orbital axes $\theta$.  
For the radial velocity of the system $v_z$, we considered independently both the blue-shifted (relative the local standard of rest) measurement of $-70$~km~s$^{-1}$  of \cite{1994MNRAS.268..430J} and the more recent measurement, which shows no evidence for a shift from local standard of rest \cite[][]{2011ApJ...732L..11N}. 
Because the parameters have likely evolved since the supernova explosion, we have used values much greater than the formal uncertainties.

  \begin{table}
\centering
\caption{ System properties $p$ and uncertainties $\delta p$ used to constrain the initial system.  \label{tab:constraints}   }
\begin{tabular}{lcc}
\hline \hline\
Parameter & $p_{\rm obs}$ &  $\Delta p$ \\  \hline
Projected semimajor axis $x$ (lt-s)  & 1296  & 45 \\ 
Orbital period $P_b$  (d) & 1237  & 124 \\
Eccentricity $e$&  0.87 & 0.4\\
L.O.S. inclination of companion spin axis $i_s$ (deg.)     &   $144$ & $4$\\
Spin-orbit misalignment $\theta$ (deg.) &  35 & 7 \\ 
Velocity in right ascension    $v_\alpha$ (km~s$^{-1}$) & $-55$ & $20$  \\
Velocity in declination  $v_\delta$ (km~s$^{-1}$) &  $20$  &  $15$\\
Radial velocity (1) $v_z$  (km~s$^{-1}$) &$70$ & $20$ \\ 
Radial velocity (2) $v_z$ (km~s$^{-1}$) & $0$    &  $20$\\ 

 \hline
 \end{tabular}
\end{table}

We searched through the parameter space using a Metropolis-Hastings algorithm Markov chain to measure the distribution of both model parameters and potentially observable properties that can be derived from the initial parameters. 

 The current orbital configuration and the system velocity constrain  the birth  properties of the system to the ranges listed in Table \ref{tab:sim_param}.  
 
We find that the mass of the pulsar progenitor  ($M_1 \approx 30~M_{\sun}$)  likely had a slightly larger mass than the mass of the $\approx 20~M_{\sun}$ companion star  
At the time of the explosion the stars were separated by $\approx 1$~AU, similar in size to the semiminor axis of the current orbit.

 We find that the pulsar was given a kick with speed $ |\bm{v}|    \approx 100$~km~s$^{-1}$, which is consistent with, but on the low end of, the distribution of pulsar velocities \cite[][]{2005MNRAS.360..974H}.  The kick velocity has values antiparallel to the spin axis of the pulsar and in the direction of the plane.
Interestingly, the kick has a significant component in the plane perpendicular to the spin of the companion. 

If the pulsar and companion spin axes are aligned, the pulsar is  a member of the $\approx 30$\% fraction of the population of young pulsars that show kicks misaligned with the spin axis \cite[][]{2005MNRAS.364.1397J,2007ApJ...664..443R,2013MNRAS.430.2281N}. 
If the kick was imparted over many rotational periods of the pulsar, the component in the plane perpendicular to the spin would  be expected to average out.  Our results suggest that any post-explosion kick would have to be imparted over at most a few rotational periods.      

As our measurements of $\ddot{x}$ and $\ddot{\omega}$ are only marginally significant we do not place strong constraints on the sky geometry of the obit. 
We find that longitude of ascending node of the orbit is $\Omega \sim -50^\circ$, which is in general agreement with an analysis of very long baseline interferometric observations presented by \cite{2011ApJ...732L..10M}.  

Best fitting parameters derived using the two different assumptions on the radial velocity yield modestly different system configurations.    If the  radial velocity is $-70$~km~s$^{-1}$, the progenitor  was likely slightly more massive, in a more compact orbit, and imparted a slightly larger kick at the time of the supernova explosion.


\begin{table*}
\centering
\caption{Plausible system initial configurations.    We placed prior constraints on the masses of the planets and the inclination angle of the planet.  For the initial inclination of the companion we assume an normal distribution $N(\mu,\sigma)$ with mean $\mu$ and standard deviation $\sigma$. The kick velocity directions are defined as follows: $v_r$ is in the direction away from the companion, $v_\phi$ is azimuthal direction, in the direction of rotation; $v_z$ is aligned with the angular momentum vector of the (pre-supernova) orbital plane \label{tab:sim_param}   }
\begin{tabular}{lllll}
\hline \hline\
Parameter & Symbol & Prior &  Best-fitting Range (1) & Best-fitting range (2)   \\
 &  & & ($v_z = 70$~km~s$^{-1}$)  & ($v_z = 0$~km~s$^{-1}$) \\ \hline
Mass of pulsar progenitor  ($M_{\sun}$) &  $M_1$ & $U(8, 50)$ &   $30 \pm 9$  & $24 \pm 9$ \\  
Mass of companion ($M_{\sun}$)  & $M_2$ &  $U(10,~30)$ & $24 \pm 4$ & $23\pm4$ \\
Orbital separation  (au)  &$A_0$& \nodata  &  $2 \pm 1$ & $6 \pm 3$ \\
Orbital phase (deg.) & $\varphi$ & \nodata &  $-40\pm 120$ & $-13 \pm 124$ \\
Kick velocity (km~s$^{-1}$) & $v_r $ & \nodata &  $80 \pm 30$ & $-5 \pm 60$   \\
Kick velocity  (km~s$^{-1}$) & $v_\phi $ & \nodata & $-70 \pm 20$ & $-55  \pm 13$     \\
Kick velocity  (km~s$^{-1}$) & $v_z $ & \nodata&  $-60 \pm 30$  & $-24 \pm 20$\\
P.A. of spin axis relative to N.C.P. (deg.) &  $\psi$  & \nodata &  $110 \pm 40$ & $100 \pm 40$\\ 
\hline 
\multicolumn{4}{c}{Derived Properties}  \\
\hline
Kick speed & $|\bm{v}|$ (km~s$^{-1}$)  & \nodata  & $130 \pm 20$  & $85 \pm 25$  \\
Longitude of ascending node (deg.)  & $\Omega$ & \nodata& $-40 \pm 90$ & $-40 \pm 100$  \\
\hline
 \end{tabular}
\end{table*}

\subsection{Orbital period increase and mass loss}

 The most likely interpretation for the $2\sigma$ measurement of $\dot{P}_b$ is mass loss from the companion due to its stellar wind.   Mass loss of rate $\dot{m}$ causes a change of the orbital period  \cite[][]{1924MNRAS..85....2J,1925MNRAS..85..912J} of
\be
&&\frac{\dot{P}_b}{P_b} = \frac{2 \dot{m}}{M_{\rm NS} + M_c}.
\ee
The mass-loss rate is therefore
\be
\dot{m} = \frac{\dot{P}_b}{2 P_b} (M_{\rm NS} + M_c)  \approx 4 \times 10^{-8} M_{\sun}~{\rm yr}^{-1},
\ee
which is consistent with the mass-loss rate of $5 \times 10^{-8}$ derived from observations of radio continuum  emission generated by the interaction of the pulsar with its companion's excretion disk \cite[][]{1996MNRAS.279.1026J}. 

There is also an apparent change in the orbital period due to a changing Doppler shift as the system crosses the sky
\cite[][]{1970SvA....13..562S,1996ApJ...467L..93K}.     This term causes an increase in $\dot{P}_b$ of 
\be
&&\Delta \dot{P}_b^{\rm PM} = \frac{|\bm{\mu}|^2 D}{c}, \nonumber \\
 &&~~~~~= 9 \times 10^{-11} \left( \frac{\mu}{12~{\rm mas~yr}^{-1}}\right)^2 \left(\frac{D}{2.3~{\rm kpc}} \right)
\ee
where  $\bm{\mu}$ is the system's proper motion, $D$ is the distance to the pulsar.  For the PSR~B1259$-$63 system this is negligible compared to the measured orbital period increase.  

\subsection{Corrections to $\dot{\omega}$ and $\dot{x}$ }\label{sec:implications_gr}

The orbit is expected to precess due to general relativistic effects and the apparent motion of the binary across the sky.    The precession rate \cite[][]{robertson1938,1986ARA&A..24..537B}  is
\be
\dot{\omega} &&= \frac{3 (G M_{\rm tot})^{2/3} n^{5/3}}{(1-e^2)c^2}. \\ \nonumber
 && = 4.4 \times 10^{-5} {\rm deg~yr}^{-1}  \left(\frac{M_{\rm tot}}{20 M_{\sun}}\right)^{2/3} ,
\ee
where $n \equiv 2\pi/P_b$ is the angular orbital frequency.  

The motion of the system across the sky causes contributes an additional term to the periastron advance \cite[][]{1996ApJ...467L..93K}:
\be
\dot{\omega}_{\rm PM} &&=  2.8 \times 10^{-7}~{\rm deg~yr}^{-1}~ \csc i_L \left(\mu_\alpha \cos \delta  \cos \Omega + \mu_\delta \sin \Omega \right)   \nonumber.
\ee
For the PSR~B1259$-$63 system, given the uncertainty in $\Omega$ can range between $ | \dot \omega_{\rm PM}| \lesssim 4 \times 10^{-6}$~deg~yr$^{-1}$   if we use an inclination angle of $i_L = 153 \pm 4 ^\circ$ (based on measurements of mass of the companion)  and  assume $\Omega$ is unconstrained. This corresponds to a less than $20\%$ correction to $\dot{\omega}$ and is approximately twice the formal measurement uncertainty in $\dot{\omega}$.  Given the uncertainty in the measurement of $\Omega$, we have not applied this correction in any of the analysis presented here. 

The system motion causes variation in the projected orbital radius to change and  $\dot{x}$ to vary   \cite[][]{1996ApJ...467L..93K}:
\be
\dot{x}_{\rm PM}  = 1.5 \times 10^{-16}~{\rm s~s}^{-1}~x \cot i_L \left(  -\mu_\alpha  \cos \delta \sin \Omega   + \mu_\delta \cos \Omega \right) 
\ee
For the PSR~B1259$-$63 system this is $ | \dot{x}_{\rm PM}|  \lesssim 3  \times 10^{-12}$~s~s$^{-1}$,  which is a factor of $10$ smaller than the measured $\dot{x}$.

\subsection{Other relativistic effects}

The Shapiro delay is large for this system, inducing a $137~\mu$s rms delay on our arrival times for a $20~M_{\sun}$ star and an inclination angle of $22^\circ$.  However,  the effect is  nearly covariant with other orbital parameters and is absorbed into the fit \cite[][]{1998MNRAS.298..997W} because of the absence of pulsed radio emission from close to periastron passage.  As mentioned in Section \ref{sec:timing}, unsurprisingly, we did not detect this Shapiro delay.   

All other relativistic effects are negligible, including the effects of relativistic orbital decay due to gravitational radiation  because of the wide separation between the two stars.

\section{Conclusions}

We have modelled 23~yr of pulse arrival-time measurements of the binary PSR~B1259$-$63 and presented a phase-connected timing solution for the observations. The solution was found after modelling the residuals with a   strong red-noise component that we attribute to rotational instabilities of the pulsar. 
In summary: \\
1)  There is strong evidence for precession of the orbital plane due to spin-orbit coupling.   We use this precession to show that the misalignment of the spin axis of the companion star LS~2883 and the orbital axis is approximately $35^\circ$.    \\
2) Combined with optical observations, we find that the companion is rotating at near-breakup velocity and has an apsidal-motion  constant of $k \approx 10^{-3}$.  \\
3)  We have measured the proper motion of the system. Combined with radial velocity measurements and distances derived from optical spectroscopy of the system,  this enables us to compute a three-dimensional velocity for the system, and find the speed of the system is $\sim 80$~to~$100$~km~s$^{-1}$ relative to the local standard of rest, with the larger values favoured for larger radial velocites.     \\ 
4)  The position, proper motion, distance, kinematic age and spindown age of the system suggest it formed in the Centaurus OB1 association.    \\
5) By combining the three-dimensional velocity of the system with the geometry of the orbit, we are able to constrain the masses  of the neutron star progenitor  and the kick the pulsar received at (or shortly after) the  explosion.   The misalignment is consistent with the pulsar receiving a kick of $\sim 80$~to~$100$~km~s$^{-1}$ at the time of  the supernova  explosion.  \\
6)  The mass of the progenitor to the pulsar was $\approx 30~M_{\sun}$. \\
7)  The binary period is increasing at a rate consistent with the companion losing mass at a rate of approximately $4 \times 10^{-8} M_{\sun}$~yr$^{-1}$.\\

\section*{Acknowledgments}

We thank the referee, N. Wex, for helpful comments that improved the quality of the manuscript and the many observers who have assisted in acquisition of the data analysed here. 
The Parkes radio telescope is part of the Australia Telescope National Facility which is funded by the Commonwealth of Australia for operation as a National Facility managed by CSIRO.   This work made use of NASA's ADS system.

\bibliographystyle{mn2ea}
\bibliography{mn-jour,thesis}

\begin{thebibliography}{47}
\expandafter\ifx\csname natexlab\endcsname\relax\def\natexlab#1{#1}\fi

\bibitem[{{Abdo} {et~al}\mbox{.}(2011){Abdo}, {Ackermann}, {Ajello},
  {Allafort}, {Ballet}, {Barbiellini}, {Bastieri}, {Bechtol}, {Bellazzini},
  {Berenji}, {Blandford}, {Bonamente}, {Borgland}, {Bregeon}, {Brigida},
  {Bruel}, {Buehler}, {Buson}, {Caliandro}, {Cameron}, {Camilo}, {Caraveo},
  {Cecchi}, {Charles}, {Chaty}, {Chekhtman}, {Chernyakova}, {Cheung}, {Chiang},
  {Ciprini}, {Claus}, {Cohen-Tanugi}, {Cominsky}, {Corbel}, {Cutini},
  {D'Ammando}, {de Angelis}, {den Hartog}, {de Palma}, {Dermer}, {Digel},
  {Silva}, {Dormody}, {Drell}, {Drlica-Wagner}, {Dubois}, {Dubus}, {Dumora},
  {Enoto}, {Espinoza}, {Favuzzi}, {Fegan}, {Ferrara}, {Focke}, {Fortin},
  {Fukazawa}, {Funk}, {Fusco}, {Gargano}, {Gasparrini}, {Gehrels}, {Germani},
  {Giglietto}, {Giommi}, {Giordano}, {Giroletti}, {Glanzman}, {Godfrey},
  {Grenier}, {Grondin}, {Grove}, {Grundstrom}, {Guiriec}, {Gwon}, {Hadasch},
  {Harding}, {Hayashida}, {Hays}, {J{\'o}hannesson}, {Johnson}, {Johnson},
  {Johnston}, {Kamae}, {Katagiri}, {Kataoka}, {Keith}, {Kerr},
  {Kn{\"o}dlseder}, {Kramer}, {Kuss}, {Lande}, {Lee}, {Lemoine-Goumard},
  {Longo}, {Loparco}, {Lovellette}, {Lubrano}, {Manchester}, {Marelli},
  {Mazziotta}, {Michelson}, {Mitthumsiri}, {Mizuno}, {Moiseev}, {Monte},
  {Monzani}, {Morselli}, {Moskalenko}, {Murgia}, {Nakamori}, {Naumann-Godo},
  {Neronov}, {Nolan}, {Norris}, {Noutsos}, {Nuss}, {Ohsugi}, {Okumura},
  {Omodei}, {Orlando}, {Paneque}, {Parent}, {Pesce-Rollins}, {Pierbattista},
  {Piron}, {Porter}, {Possenti}, {Rain{\`o}}, {Rando}, {Ray}, {Razzano},
  {Razzaque}, {Reimer}, {Reimer}, {Reposeur}, {Ritz}, {Sadrozinski}, {Scargle},
  {Sgr{\`o}}, {Shannon}, {Siskind}, {Smith}, {Spandre}, {Spinelli},
  {Strickman}, {Suson}, {Takahashi}, {Tanaka}, {Thayer}, {Thayer}, {Thompson},
  {Thorsett}, {Tibaldo}, {Tibolla}, {Torres}, {Tosti}, {Troja}, {Uchiyama},
  {Usher}, {Vandenbroucke}, {Vasileiou}, {Vianello}, {Vitale}, {Waite}, {Wang},
  {Winer}, {Wolff}, {Wood}, {Wood}, {Yang}, {Ziegler}, \&
  {Zimmer}}]{2011ApJ...736L..11A}
{Abdo} A.~A. {et~al.}, 2011, \apjl, 736, L11

\bibitem[{Akaike(1974)}]{1100705}
Akaike H., 1974, Automatic Control, IEEE Transactions on, 19, 716

\bibitem[{{Backer} \& {Hellings}(1986)}]{1986ARA&A..24..537B}
{Backer} D.~C., {Hellings} R.~W., 1986, \araa, 24, 537

\bibitem[{{Brandt} \& {Podsiadlowski}(1995)}]{1995MNRAS.274..461B}
{Brandt} N., {Podsiadlowski} P., 1995, \mnras, 274, 461

\bibitem[{{Claret}(1999)}]{1999A&A...350...56C}
{Claret} A., 1999, \aap, 350, 56

\bibitem[{{Coles} {et~al}\mbox{.}(2011){Coles}, {Hobbs}, {Champion},
  {Manchester}, \& {Verbiest}}]{2011MNRAS.418..561C}
{Coles} W., {Hobbs} G., {Champion} D.~J., {Manchester} R.~N., {Verbiest}
  J.~P.~W., 2011, \mnras, 418, 561

\bibitem[{{Connors} {et~al}\mbox{.}(2002){Connors}, {Johnston}, {Manchester},
  \& {McConnell}}]{2002MNRAS.336.1201C}
{Connors} T.~W., {Johnston} S., {Manchester} R.~N., {McConnell} D., 2002,
  \mnras, 336, 1201

\bibitem[{{Cordes} \& {Shannon}(2010)}]{cs2010}
{Cordes} J.~M., {Shannon} R.~M., 2010, ArXiv:1010.3785

\bibitem[{{Corti} \& {Orellana}(2013)}]{2013A&A...553A.108C}
{Corti} M.~A., {Orellana} R.~B., 2013, \aap, 553, A108

\bibitem[{{Edwards}, {Hobbs} \& {Manchester}(2006){Edwards}, {Hobbs}, \&
  {Manchester}}]{2006MNRAS.372.1549E}
{Edwards} R.~T., {Hobbs} G.~B., {Manchester} R.~N., 2006, \mnras, 372, 1549

\bibitem[{{Everett} \& {Weisberg}(2001)}]{2001ApJ...553..341E}
{Everett} J.~E., {Weisberg} J.~M., 2001, \apj, 553, 341

\bibitem[{{Hobbs} {et~al}\mbox{.}(2005){Hobbs}, {Lorimer}, {Lyne}, \&
  {Kramer}}]{2005MNRAS.360..974H}
{Hobbs} G., {Lorimer} D.~R., {Lyne} A.~G., {Kramer} M., 2005, \mnras, 360, 974

\bibitem[{{Jeans}(1924)}]{1924MNRAS..85....2J}
{Jeans} J.~H., 1924, \mnras, 85, 2

\bibitem[{{Jeans}(1925)}]{1925MNRAS..85..912J}
{Jeans} J.~H., 1925, \mnras, 85, 912

\bibitem[{{Johnston} {et~al}\mbox{.}(2005{\natexlab{a}}){Johnston}, {Ball},
  {Wang}, \& {Manchester}}]{2005MNRAS.358.1069J}
{Johnston} S., {Ball} L., {Wang} N., {Manchester} R.~N., 2005{\natexlab{a}},
  \mnras, 358, 1069

\bibitem[{{Johnston} {et~al}\mbox{.}(2005{\natexlab{b}}){Johnston}, {Hobbs},
  {Vigeland}, {Kramer}, {Weisberg}, \& {Lyne}}]{2005MNRAS.364.1397J}
{Johnston} S., {Hobbs} G., {Vigeland} S., {Kramer} M., {Weisberg} J.~M., {Lyne}
  A.~G., 2005{\natexlab{b}}, \mnras, 364, 1397

\bibitem[{{Johnston} {et~al}\mbox{.}(1992){Johnston}, {Manchester}, {Lyne},
  {Bailes}, {Kaspi}, {Qiao}, \& {D'Amico}}]{1992ApJ...387L..37J}
{Johnston} S., {Manchester} R.~N., {Lyne} A.~G., {Bailes} M., {Kaspi} V.~M.,
  {Qiao} G., {D'Amico} N., 1992, \apjl, 387, L37

\bibitem[{{Johnston} {et~al}\mbox{.}(1996){Johnston}, {Manchester}, {Lyne},
  {D'Amico}, {Bailes}, {Gaensler}, \& {Nicastro}}]{1996MNRAS.279.1026J}
{Johnston} S., {Manchester} R.~N., {Lyne} A.~G., {D'Amico} N., {Bailes} M.,
  {Gaensler} B.~M., {Nicastro} L., 1996, \mnras, 279, 1026

\bibitem[{{Johnston} {et~al}\mbox{.}(1994){Johnston}, {Manchester}, {Lyne},
  {Nicastro}, \& {Spyromilio}}]{1994MNRAS.268..430J}
{Johnston} S., {Manchester} R.~N., {Lyne} A.~G., {Nicastro} L., {Spyromilio}
  J., 1994, \mnras, 268, 430

\bibitem[{{Kaspi} {et~al}\mbox{.}(1996){Kaspi}, {Bailes}, {Manchester},
  {Stappers}, \& {Bell}}]{1996Natur.381..584K}
{Kaspi} V.~M., {Bailes} M., {Manchester} R.~N., {Stappers} B.~W., {Bell} J.~F.,
  1996, \nat, 381, 584

\bibitem[{{Kopeikin}(1995)}]{1995ApJ...439L...5K}
{Kopeikin} S.~M., 1995, \apjl, 439, L5

\bibitem[{{Kopeikin}(1996)}]{1996ApJ...467L..93K}
{Kopeikin} S.~M., 1996, \apjl, 467, L93

\bibitem[{{Lai}(2004)}]{2004cetd.conf..276L}
{Lai} D., 2004, in Cosmic explosions in three dimensions, {H{\"o}flich} P.,
  {Kumar} P., {Wheeler} J.~C., eds., p. 276

\bibitem[{{Lambert} \& {Rickett}(1999)}]{1999ApJ...517..299L}
{Lambert} H.~C., {Rickett} B.~J., 1999, \apj, 517, 299

\bibitem[{{Manchester} \& {Johnston}(1995)}]{1995ApJ...441L..65M}
{Manchester} R.~N., {Johnston} S., 1995, \apjl, 441, L65

\bibitem[{{Manchester} {et~al}\mbox{.}(1995){Manchester}, {Johnston}, {Lyne},
  {D'Amico}, {Bailes}, \& {Nicastro}}]{1995ApJ...445L.137M}
{Manchester} R.~N., {Johnston} S., {Lyne} A.~G., {D'Amico} N., {Bailes} M.,
  {Nicastro} L., 1995, \apjl, 445, L137

\bibitem[{{McClure-Griffiths} {et~al}\mbox{.}(1998){McClure-Griffiths},
  {Johnston}, {Stinebring}, \& {Nicastro}}]{1998ApJ...492L..49M}
{McClure-Griffiths} N.~M., {Johnston} S., {Stinebring} D.~R., {Nicastro} L.,
  1998, \apjl, 492, L49

\bibitem[{{Melatos}, {Johnston} \& {Melrose}(1995){Melatos}, {Johnston}, \&
  {Melrose}}]{1995MNRAS.275..381M}
{Melatos} A., {Johnston} S., {Melrose} D.~B., 1995, \mnras, 275, 381

\bibitem[{{Mold{\'o}n} {et~al}\mbox{.}(2011){Mold{\'o}n}, {Johnston},
  {Rib{\'o}}, {Paredes}, \& {Deller}}]{2011ApJ...732L..10M}
{Mold{\'o}n} J., {Johnston} S., {Rib{\'o}} M., {Paredes} J.~M., {Deller} A.~T.,
  2011, \apjl, 732, L10

\bibitem[{{Murray} \& {Dermott}(1999)}]{1999ssd..book.....M}
{Murray} C.~D., {Dermott} S.~F., 1999, {Solar system dynamics}. Cambridge
  University Press

\bibitem[{{Negueruela} {et~al}\mbox{.}(2011){Negueruela}, {Rib{\'o}},
  {Herrero}, {Lorenzo}, {Khangulyan}, \& {Aharonian}}]{2011ApJ...732L..11N}
{Negueruela} I., {Rib{\'o}} M., {Herrero} A., {Lorenzo} J., {Khangulyan} D.,
  {Aharonian} F.~A., 2011, \apjl, 732, L11

\bibitem[{{Noutsos} {et~al}\mbox{.}(2013){Noutsos}, {Schnitzeler}, {Keane},
  {Kramer}, \& {Johnston}}]{2013MNRAS.430.2281N}
{Noutsos} A., {Schnitzeler} D.~H.~F.~M., {Keane} E.~F., {Kramer} M., {Johnston}
  S., 2013, \mnras, 430, 2281

\bibitem[{{Radhakrishnan} \& {Cooke}(1969)}]{1969ApL.....3..225R}
{Radhakrishnan} V., {Cooke} D.~J., 1969, \aplett, 3, 225

\bibitem[{{Rankin}(2007)}]{2007ApJ...664..443R}
{Rankin} J.~M., 2007, \apj, 664, 443

\bibitem[{Robertson(1938)}]{robertson1938}
Robertson H.~P., 1938, Annals of Mathematics, 39, 101

\bibitem[{{Shannon} \& {Cordes}(2010)}]{sc2010}
{Shannon} R.~M., {Cordes} J.~M., 2010, \apj, 725, 1607

\bibitem[{{Shapiro}(1964)}]{1964PhRvL..13..789S}
{Shapiro} I.~I., 1964, Physical Review Letters, 13, 789

\bibitem[{{Shklovskii}(1970)}]{1970SvA....13..562S}
{Shklovskii} I.~S., 1970, Sov. Ast., 13, 562

\bibitem[{{Spruit} \& {Phinney}(1998)}]{1998Natur.393..139S}
{Spruit} H., {Phinney} E.~S., 1998, \nat, 393, 139

\bibitem[{{Taylor}(1992)}]{1992RSPTA.341..117T}
{Taylor} J.~H., 1992, Royal Society of London Philosophical Transactions Series
  A, 341, 117

\bibitem[{{Tetzlaff}, {Neuh{\"a}user} \& {Hohle}(2011){Tetzlaff},
  {Neuh{\"a}user}, \& {Hohle}}]{2011MNRAS.410..190T}
{Tetzlaff} N., {Neuh{\"a}user} R., {Hohle} M.~M., 2011, \mnras, 410, 190

\bibitem[{{van den Heuvel}(1984)}]{1984JApA....5..209V}
{van den Heuvel} E.~P.~J., 1984, Journal of Astrophysics and Astronomy, 5, 209

\bibitem[{{Wang}, {Lai} \& {Han}(2006){Wang}, {Lai}, \&
  {Han}}]{2006ApJ...639.1007W}
{Wang} C., {Lai} D., {Han} J.~L., 2006, \apj, 639, 1007

\bibitem[{{Wang}, {Johnston} \& {Manchester}(2004){Wang}, {Johnston}, \&
  {Manchester}}]{2004MNRAS.351..599W}
{Wang} N., {Johnston} S., {Manchester} R.~N., 2004, \mnras, 351, 599

\bibitem[{{Weltevrede} {et~al}\mbox{.}(2010){Weltevrede}, {Johnston},
  {Manchester}, {Bhat}, {Burgay}, {Champion}, {Hobbs}, {K{\i}z{\i}ltan},
  {Keith}, {Possenti}, {Reynolds}, \& {Watters}}]{2010PASA...27...64W}
{Weltevrede} P. {et~al.}, 2010, PASA, 27, 64

\bibitem[{{Wex}(1998)}]{1998MNRAS.298...67W}
{Wex} N., 1998, \mnras, 298, 67

\bibitem[{{Wex} {et~al}\mbox{.}(1998){Wex}, {Johnston}, {Manchester}, {Lyne},
  {Stappers}, \& {Bailes}}]{1998MNRAS.298..997W}
{Wex} N., {Johnston} S., {Manchester} R.~N., {Lyne} A.~G., {Stappers} B.~W.,
  {Bailes} M., 1998, \mnras, 298, 997

\end{thebibliography}
\appendix

\end{document}